\documentclass[aps,prd,superscriptaddress,showpacs,tighten,nofootinbib]{revtex4-1}
\usepackage{epsfig}
\usepackage{amssymb}
\usepackage{amsmath}
\usepackage{rotate}
\usepackage{float}
\usepackage{graphicx}
\usepackage{subfig}
\usepackage{slashed}
\newcommand{\ba}{\begin{array}}
\newcommand{\ea}{\end{array}}
\newcommand{\bd}{\begin{displaymath}}
\newcommand{\ed}{\end{displaymath}}
\newcommand{\be}{\begin{equation}}
\newcommand{\ee}{\end{equation}}
\newcommand{\bea}{\begin{eqnarray}}
\newcommand{\eea}{\end{eqnarray}}

\def\a{\alpha}

\def\th13 {\theta_{13}}

\usepackage{graphicx}

\newcommand{\degree}{\ensuremath{^\circ}}

\catcode`\@=11
\def\lsim{\mathrel{\mathpalette\@versim<}}
\def\gsim{\mathrel{\mathpalette\@versim>}}
\def\@versim#1#2{\vcenter{\offinterlineskip
\ialign{$\m@th#1\hfil##\hfil$\crcr#2\crcr\sim\crcr } }}
\catcode`\@=12

\catcode`@=12
\begin{document}
\title{Discovering $CP$ violation in neutrino oscillation experiment using \\ neutrino beam from electron capture}

\author{Zini Rahman}
\email{zini@ctp-jamia.res.in}
\affiliation{Centre for Theoretical Physics, Jamia Millia Islamia (Central University), \\ Jamia Nagar, New Delhi-110025, INDIA}

\author{Rathin Adhikari}
\email{rathin@ctp-jamia.res.in}
\affiliation{Centre for Theoretical Physics, Jamia Millia Islamia (Central University), \\ Jamia Nagar, New Delhi-110025, INDIA}

\begin{abstract}
Considering the recently obtained value of $\theta_{13}$  from Daya Bay and other reactor experiments  we have studied the 
prospects of considering mono-energetic neutrino beam in studying $CP$ violation in the leptonic sector.  Using a neutrino beam from electron capture process for nuclei $^{110}_{50}$Sn and $^{152}$Yb and considering two baselines - 130 Km and 250 Km with Water Cherenkov detector,  we have shown the discovery reach of $CP$ violation in neutrino oscillation experiment.  Particularly for $^{110}_{50}$Sn nuclei $CP$  violation could be found for about 80\%  of the possible $\delta$ values for a baseline of 130 km with boost factor $\gamma = 500$. This result is obtained with  conservative choice of neutrino energy resolution using the possible vertex resolution at the detector and taking into account beam spreading. The nuclei  $^{152}$Yb is although more suitable technically for the production of mono-energetic beam, but  is found to be not so suitable for good discovery reach of $CP$ violation. 
\end{abstract}
\maketitle 
\section{Introduction}
Long back $CP$ violation  has been found in the quark sector of the Standard Model of Particle Physics. But so far there is no evidence of $CP$ violation in the leptonic sector. One way to search for such $CP$ violation is through neutrino oscillation experiments in which one flavor of neutrino could oscillate to other flavors of neutrinos. Neutrino oscillation probability depends on various oscillation parameters present in the mixing matrix - the PMNS matrix \cite{pmns} and the neutrino mass squared differences. Two of the three angles - $\theta_{12}$ and $\theta_{23}$ present in PMNS matrix have been known with certain accuracy for some time. Recently  several experiments like Double Chooz, Daya Bay and RENO collaboration \cite{double,daya,reno} found non-zero value of $\sin^2 2 \theta_{13}$ corresponding to the third mixing angle of PMNS matrix with a global level of significance which is well above conventional  $5\sigma$ discovery threshold. As three  angles are non-zero  there could be
non-zero $CP$ violating phase $\delta $ in the PMNS matrix and as such $CP$ violation in the leptonic sector.  The mass squared differences - $|\Delta m^2_{31}|$ and $\Delta m^2_{21}$ is known (where $\Delta m^2_{ij} = m_i^2 -m_j^2$) but the sign of $\Delta m^2_{31}$ and as such the hierarchy (whether it is  normal (NH) or inverted (IH)) of neutrino masses is still unknown. Also the $CP$ violating phase $\delta$ is still unknown.  Various neutrino oscillation experiments like superbeam, neutrino factory, beta beams and reactor experiments are focussing on determining these unknown parameters corresponding to neutrino oscillations.
 
In recent years oscillation experiment using a neutrino beam with neutrinos emitted from an electron capture process is proposed \cite{rolinec,mon,mono2,mono3,mono9,mono11,mono12}. Such  beam can be produced using an accelerated nuclei that decay by electron capture. In this process an electron is captured by a proton releasing a neutron and an electron neutrino. So
the beam is purely of one flavor. In the rest frame of the mother nuclei the electron neutrino that is released from such process, has a definite energy Q. Since the idea of using a neutrino beam emitted from an electron capture process is based on the acceleration and storage of radioactive isotopes that decays to daughter nuclei, one may get the suitable neutrino energy by accelerating the mother nuclei with suitable Lorentz boost factor $\gamma$. One can control the neutrino energy by choosing the appropriate Lorentz boost factor  as the energy that has been boosted by an  appropriate boost factor towards the detector is given as $E=2 \gamma Q$. Hence for certain mother nuclei to get the required neutrino energy the  boost factor have to be chosen appropriately with respect to Q.  Due to the almost monoenergetic nature of  such beam  one can appropriately choose the neutrino energy for which the probability of oscillation could be large and sensitive to certain unknown  neutrino oscillation parameters.
In this work we consider such a flavor pure electron neutrino beam emitted from electron capture process for a suitable $\gamma$ value where the beam is targeted towards a Water Cherenkov detector and perform numerical simulation to study the discovery reach of $CP$ violation in oscillation experiments.

In section II, we discuss about two different nuclei, $^{110}_{50}$Sn and $^{152}$Yb 
which is  considered for the electron capture experiment. We also discuss our procedure
of numerical simulation.   
We show how  $\nu_e \rightarrow \nu_\mu$ oscillation probability depends on $\delta$. The
expression of probability has been obtained using perturbation method suitable for shorter baselines of length like 130 Km or 250 Km (as considered later). 
We  also discuss  procedure
for choosing suitable boost factor $\gamma $ for specific  baseline and specific  nuclei considered for the neutrino beam. In section III, we mention four experimental setups that we have considered for analysis of discovery reach of $CP$ violation with monoenergetic neutrino beam. Also for comparison we consider another experimental set-up with superbeam facility. 
Also, we mention   values of various oscillation parameters and  detector characteristics which have been considered in our simulation method. In section IV, we discuss the possible  discovery reach of $CP$ violation 
for different experimental set-ups considered for two different nuclei, $^{110}_{50}$Sn and $^{152}$Yb for different baselines. In section V, we have summarized our results and  also have discussed the difficulties and challenges in getting the monoenergetic neutrino beam.

\section{Suitable boost factor, neutrino energy from $\nu_e \rightarrow \nu_\mu$ oscillation probability}
The most suitable candidate for producing neutrino beam from electron capture process would be the one with a low Q value and high boost factor, $\gamma$ \cite{mono3}. Also it would be preferable if the nuclei has a short half life. The reason for these is as follows. We need neutrino energy around the  peak of the oscillation probability where variation due to $\delta $ is significant and as such $\frac{\Delta m^2_{31}L}{4E} \approx (2 n + 1) \frac{\pi}{2}$. Considering $E=2\gamma Q$, it follows that $\gamma = \frac{\Delta m^2_{31}L}{4\pi Q}$ as for example, for the first oscillation peak. For sufficiently high $\gamma$ almost all neutrinos are expected to go through the detector. Then to satisfy the above condition we   need to lower Q value. Then another condition is,  $\gamma \tau < T$ where $T$ is the time considered to perform the experiment so that all the nuclei decay and
$\tau $ is the half life of the nuclei. If $\gamma$ is increased then the  half life $\tau$ is required to be small. So the  preferable factors considered in choosing the candidate for producing neutrino beam from electron capture process are -  low Q value,  small half life $\tau$ and  high $\gamma$ \cite{mono3}. Although higher $\gamma $ needs technological advancement for the accelerator.

The isotope, $^{110}_{50}$Sn, has
Q = 267 KeV in the rest frame and a half life of 4.11 h. As it
has a low Q value so one may consider high $\gamma$ value. However, it has a longer half life as compared to other nuclei like $^{150}Er$, $^{152}$Yb, $^{156}$Yb, $^{150}$Dy, $^{148}$Dy \cite{mono2,mono11} whose half lifes are 18.5 seconds, 3.04 seconds, 261 seconds, 7.2 min and 3.1 min respectively. However, these nuclei have larger Q values of the order of $10^3$ KeV. On the other hand, considering effective running time
per year as $10^7$  second all the nuclei for isotope $^{110}_{50}$Sn will not decay. But for $\gamma =500$ or 320  (as considered in our analysis to  obtain the suitable  neutrino energy $E$ 
resulting in high oscillation probability)  the 
useful decays are respectively about 0.608 or 0.768 times the total number of $^{110}_{50}$Sn nuclei considered. So there is not much suppression in numbers of nuclei. Hence although $^{110}_{50}$Sn has a larger half life, due to its lower Q value there is scope to consider higher $\gamma$
for 130 Km or 250 Km baselines. For these reasons we have preferred isotope, $^{110}_{50}$Sn in comparison to other nuclei. However, there is recent study on finding suitable candidate nuclei for electron capture process and it has been found that $^{150}Er$, $^{152}$Yb, $^{156}$Yb nuclei have dominant electron capture decay
 to one level. Particularly, $^{152}$Yb has been found to be most suitable one \cite{mono11}.
For this reason,  apart from nuclei $^{110}_{50}$Sn we shall consider $^{152}$Yb also for our analysis. 
 However, as $Q$ value (5435 KeV) for $^{152}$Yb is higher,  corresponding $\gamma$ value for such nuclei are supposed to be small.

The  neutrino beam produced from electron capture process is boosted with certain boost factor, $\gamma$. The boosted neutrino beam produced from such process hits the detector at a baseline of length  $L$ at a radial distance $R$ from the beam axis and the energy, $E$ of this beam in rest frame of the detector , i.e., in laboratory frame is given by:

\begin{eqnarray}
\label{eq:erest}
E  (R)=\frac{Q}{\gamma}\bigg[1-\frac{\beta}{\sqrt{1+(R/L)^2}}\bigg]^{-1}\approx \frac{2\gamma Q}{1+(\gamma R/L)^2}
\end{eqnarray}
where R is the radial distance at the detector from the beam axis. At beam center, $R=0$. From the above equation (\ref{eq:erest}) the energy window considered for the analysis which is constrained by the size of the detector is given by:

\begin{equation}
\label{eq:ewindow}
\frac{2\gamma Q}{1+(\gamma R _{max}/L)^2} \leq E \leq 2\gamma Q
\end{equation}
 From equation (\ref{eq:ewindow}) we can see that once the baseline length L and $\gamma$ is fixed the energy window gets fixed.  
However, even considering  radius of the detector  $R_{max}=100 $ m the energy window is very small as can be seen from figure 1.

One can see from equation (1) that it is possible to tell precisely the energy from the $R$ value of the Cherenkov ring at the Cherenkov detector instead of measuring directly the neutrino energy. So there is scope to get good energy resolution by measuring position if the vertex resolution is good. The $\sigma(E)$ function corresponding to energy resolution function (as used in running GLoBES \cite{globes1})
 in terms of vertex measurement uncertainty $\sigma(R)$ can be written as:
\bea
\sigma(E)=-\frac{Q R \beta }{L^2 \left(1+\frac{R^2}{L^2}\right)^{3/2} \left(1-\frac{\beta }{\sqrt{1+\frac{R^2}{L^2}}}\right)^2 \gamma }\sigma(R)
\label{eq:ereso}
\eea
where $\beta$ is defined as
\bea
\beta=\frac{\sqrt{\gamma ^2-1}}{\gamma }
\label{eq:beta}
\eea
Vertex measurement uncertainty for electron/muon identification at Super-K is around 30 cm \cite{skvertex}. To estimate  the energy resolution using position measurement one may consider $\sigma(R) \sim 30 $ cm provided that the beam spreading $\frac{\sigma(R)}{L}$ is negligible ( lesser than
about $1 \mu$rad) \cite{rolinec} which is difficult to achieve experimentally.
 If we take into  account the beam divergence about 10 $\mu rad$  (which is 
almost one order larger than that considered in references \cite{ beam, divergence}), one may consider larger $\sigma(R) \sim 130 $ cm    particularly for baseline of 130 Km. For baseline like 
250 Km it would be more but we have considered same $\sigma(R)$ which means the beam divergence has been assumed to be lesser than about 5 $\mu rad$ for the analysis for  baseline with length 250 Km. 

In this work GLoBES\cite{globes1}  has been used for doing the simulations.
In order to use this software, the radial binning is replaced by binning in energy and the bins are not equidistant.
If we divide $R^2_{max}$ into $k$ bins  the edges of the bins are given as:
\begin{eqnarray}
\label{eq:r}
R^2_i=R^2_{max}-(i-1)\Delta R^2 
\end{eqnarray}
with
\begin{eqnarray}
\label{eq:dr}
\Delta R^2=\frac{R^2_{max}}{k} 
\end{eqnarray}
We consider $R^2_i>R^2_{i+1}$ so that in GLoBES the respective energy bins are in the correct order as given below
\begin{eqnarray} 
\label{eq:ep}
E(R^2_i)<E(R^2_{i+1})
\end{eqnarray}
where 
$E$ is the neutrino energy in the lab frame.

The number of events per bin $i$ and channel $c$ (different channels mentioned later in this section)  is given by:
\begin{eqnarray}
\label{eq:event}
N_{event} \simeq \frac{N_{norm}}{L^2} \int_{E_i - \Delta E_i/2}^{E_i + \Delta E_i/2} dE' \int_0^\infty
dE \phi (E) P^c(L,E) \sigma^c(E)  \epsilon^c(E') R^c(E,E')
\end{eqnarray}
where $N_{norm}$ is the normalization factor for using GLoBES and is related to length of the baseline, area and energy binning related to 
flux , number of target nuclei per unit target mass and number of nuclei decaying.  
$\epsilon^c $ is the signal efficiency in the respective channel,   $P^c(L,E)$ is the neutrino oscillation probability in particular channel, $\sigma^c(E)$ is total cross section for particular flavor of neutrinos and particular interaction corresponding to particular channel. $R(E,E')$ is the energy resolution function of the detector where $E'$ is the reconstructed neutrino energy. $\phi (E)$ has been calculated from  the angular neutrino flux 
$\frac{dn}{d\Omega}(E)$ as defined below :
\bea
\label{eq:flux}
\frac{dn}{d\Omega}=\frac{N_{decays}}{4\pi}\left(\frac{E}{Q}\right)^2
\eea
where $N_{decays}$ is number of nuclei actually decaying per year. The detailed derivation of these expressions can be found in \cite{rolinec}. Considering equation (\ref{eq:ewindow})
and ( \ref{eq:flux}) one can see that with increase in $\gamma$ value the angular flux increases.  

Out of      $\nu_\mu$ appearance channel and $\nu_e$ disappearance channels considered 
in our analysis, in the  appearance channel, i.e,  ${\nu_e \rightarrow \nu_{\mu}}$ oscillation channel, 
$CP$ violating phase $\delta $ appears at the order of $\alpha^{3/2}$ (where $\alpha= \Delta m_{21}^2/\Delta m_{31}^2$) . The effect of $\delta $ in the probability of other oscillation channels are lesser. Following the perturbation method \cite{ra} and by considering the standard model matter effect $A \sim \alpha$  for neutrino energy $E$ around 1 GeV and  based on recent Daya Bay and other experimental results  $\sin \theta_{13} \sim \sqrt{\alpha} $,  
 $A = 2 \sqrt{2} G_F n_e E/\Delta m_{31}^2 $  and $n_e$ is the electron number density in matter; the probability of oscillation $P_{\nu_e \rightarrow \nu_{\mu}}$ for short baseline (of 130 Km or 250 Km as considered in the analysis) upto order $\alpha^{2}$ is given by

\begin{widetext}
 \begin{eqnarray}
  P_{\nu_e \rightarrow \nu_{\mu}} = P_{\nu_e \rightarrow \nu_{\mu}}(\alpha) + P_{\nu_e \rightarrow \nu_{\mu}}(\alpha^{3/2}) + P_{\nu_e \rightarrow \nu_{\mu}}(\alpha^2) 
\label{eq:p250}
\end{eqnarray}
\end{widetext}
where 
\begin{widetext}
 \begin{eqnarray}
 P_{\nu_e \rightarrow \nu_{\mu}}(\alpha)&=& \sin^2 2\theta_{13}\sin^2 \theta_{23}\sin^2\bigg(\frac{\Delta m^2_{31}L}{4E}\bigg) \nonumber \\
  P_{\nu_e \rightarrow \nu_{\mu}}(\alpha^{3/2}) &= &\alpha\bigg(\frac{\Delta m^2_{31}L}{2E}\bigg) \sin \theta_{13}\sin 2\theta_{23}\sin 2 \theta_{12}\sin \bigg(\frac{\Delta m^2_{31}L}{4E}\bigg)
 \cos \bigg(\delta - \frac{\Delta m^2_{31}L}{4E}\bigg) \nonumber \\
 P_{\nu_e \rightarrow \nu_{\mu}}(\alpha^2)&=&  \a^2 \cos^2\theta_{23}\sin^2 2\theta_{12}\bigg(\frac{\Delta m^2_{31}L }{4E}\bigg)^2-2\a \sin^2 \theta_{13}\sin^2 \theta_{12}\sin^2 \theta_{23}\frac{\Delta m^2_{31}L}{2E}
 \sin \frac{\Delta m ^2_{31}L}{2E}  \nonumber \\ &+& 8A\sin^2 \theta_{13}\sin^2 \theta_{23}\bigg(\sin^2 \frac{\Delta m ^2_{31}L}{4E}-\frac{\Delta m ^2_{31}L}{8E} \sin \frac{\Delta m ^2_{31}L}{2E}\bigg)
\label{eq:p250}
\end{eqnarray}
\end{widetext}
 One may note that this expression of oscillation probability is little bit different  from that presented by earlier authors \cite{akh} because they have considered the perturbative approach for relatively longer baseline and  small $\sin\theta_{13} \sim \alpha$. Particularly the $\delta$ dependence appears to be more and instead of being at the order of $\alpha^2$ it appears at the order $\alpha^{3/2}$ although the expression is same. However at order $\alpha^2$ some extra terms appear in comparison to that presented by earlier authors although their contribution is relatively smaller being at the order of $\alpha^2$.
The matter effect  is occurring at order $\a^2$ through term containing parameter $A$. But for longer baselines matter effect increases. So
in general, the shorter baselines are preferred to find any $CP$ violating effect
due to $\delta$.
In future such $CP$ violation in the leptonic sector could be explored in superbeam, beta beam, neutrino factory 
or in mono-energetic neutrino beam (through electron capture process). However, one advantage 
of monoenergetic neutrino beam which we consider here, is that there is scope
 of setting the neutrino energy around the  probability oscillation peak which is more sensitive to any variation in $CP$ violating phase $\delta$.

\begin{figure}[H]
\centering
\begin{tabular}{cc}
\includegraphics[width=0.4\textwidth]{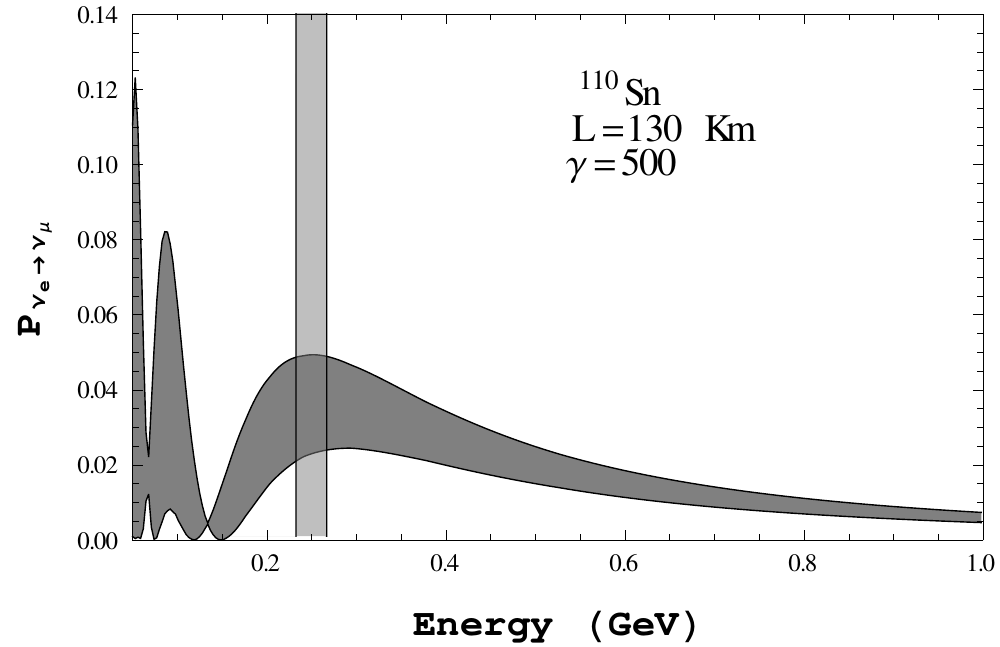}&
\includegraphics[width=0.4\textwidth]{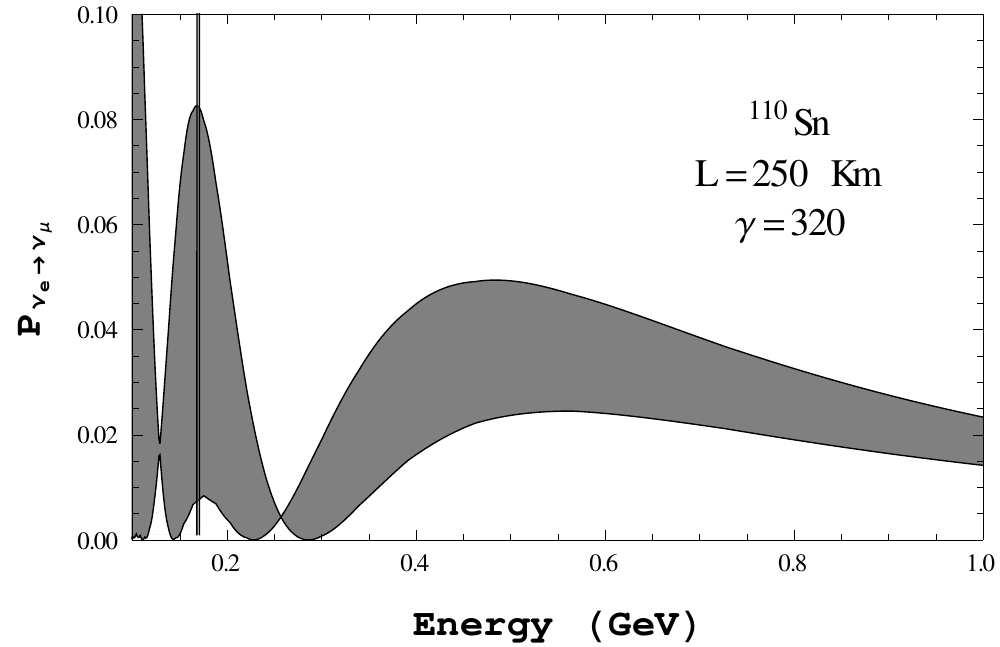}\\
\includegraphics[width=0.4\textwidth]{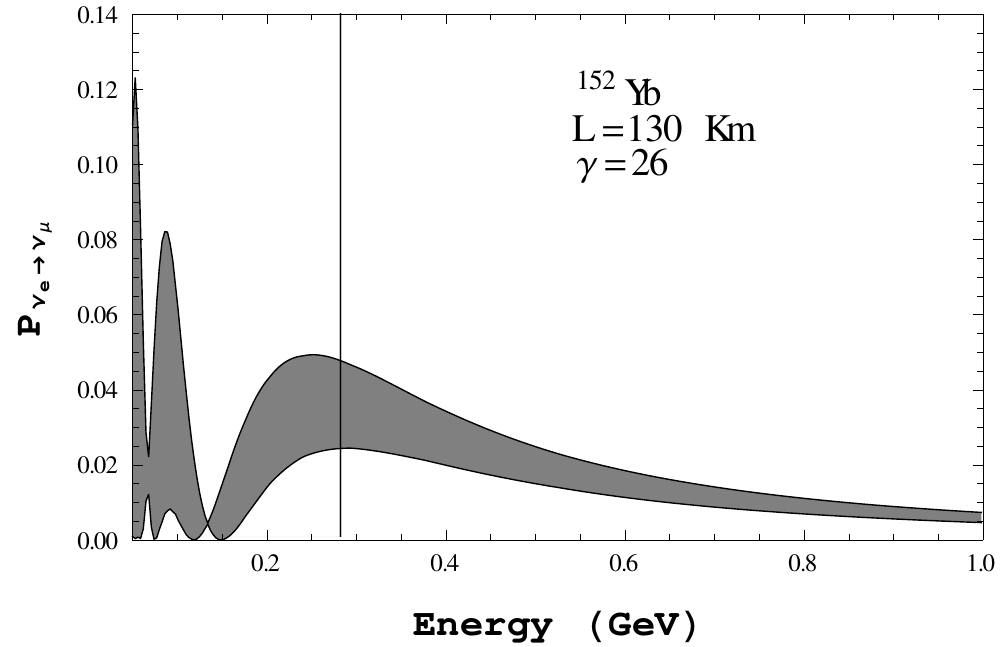}&
\includegraphics[width=0.4\textwidth]{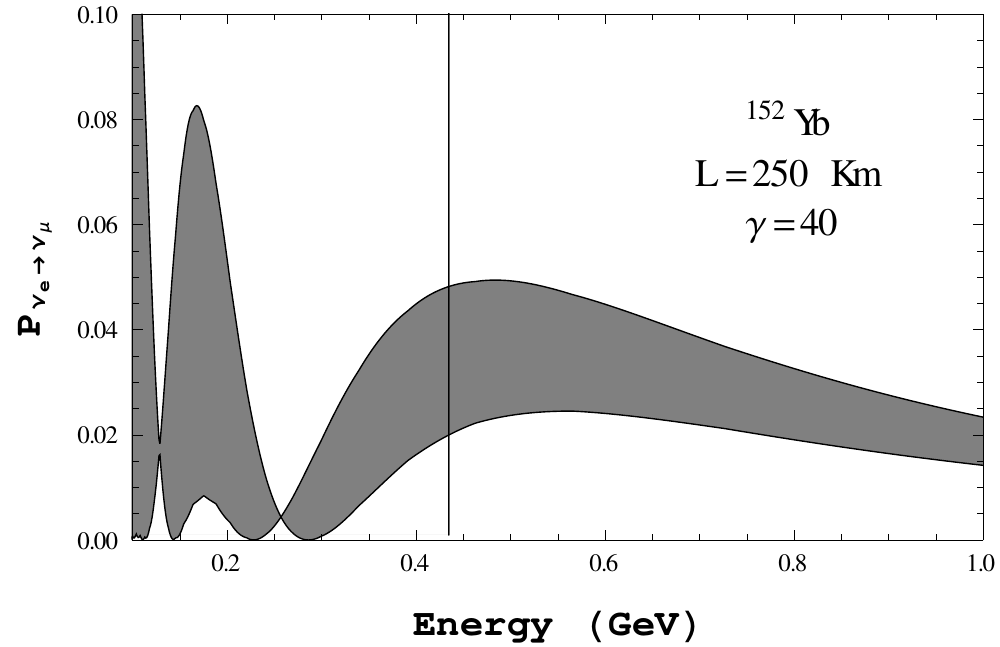}
\end{tabular}
\caption[] {{\small Probability $P(\nu_e\rightarrow \nu_{\mu})$ vs neutrino energy $E$ for 
two different nuclei $^{110}_{50}$Sn and $^{152}$Yb and the corresponding energy window satisfying equation (\ref{eq:ewindow}) for different $\gamma$ values.}} 
\label{fig1}
\end{figure}

We have plotted numerically the probability $P(\nu_{e}\rightarrow \nu_{\mu})$ with respect to energy for two different baselines of length 130 Km (CERN-Frejus) and 250 Km  for
two different nuclei $^{110}_{50}$Sn and $^{152}$Yb. We have considered normal hierarchy in plotting figure \ref{fig1}. For all plots in figure \ref{fig1},  $\delta $ has been varied over its' entire allowed range (0 to 2 $\pi$) resulting in the shaded region in each plots 
showing the significant variation of probability at particular neutrino energies. Corresponding to each of
the nuclei (whose $Q$ values are fixed) we have considered appropriate $\gamma$ value  
so that the corresponding energy window (as mentioned in (\ref{eq:ewindow})) overlaps with
the shaded region near the oscillation peaks having significant variation of probability due to $\delta$ variation. In choosing $\gamma$, one also has to keep in mind that the neutrino energy is not too low as otherwise flux will be much lesser. The energy window has been shown by the shaded vertical strips. For our suitable choice of $\gamma$ value, the energy window  is larger for 130 Km baseline and relatively smaller for 250 Km baseline for both the nuclei. Also the energy window for $^{110}_{50}$Sn is larger than  $^{152}$Yb.

 For finding $\delta$ we shall prefer  the maximum variation of the probability with $\delta$ which will occur for neutrino energy satisfying $\frac{\Delta m ^2_{31}L}{4E} \approx (2n+1) \pi/2$ where $n$ is an integer. This has been shown in  figure \ref{fig1} in which the oscillation probability has been evaluated numerically considering the evolution of neutrino flavor states.  However, the energy also depends on the $Q$ value of the corresponding nuclei. So we have considered the case of two nuclei separately. As for example, for $^{110}_{50}$Sn nuclei, for baseline of length 130 Km we have considered first oscillation maximum and for baseline of length 250 Km we have considered second oscillation maximum where the variation of the probability due to $\delta $ is significant. For $^{152}$Yb, for both the baselines we have considered first oscillation peak.  
 In considering the suitable peak in the oscillation probability one has to keep in mind that the neutrino flux varies with $E^2$ 
as shown in \eqref{eq:flux} and so after doing the numerical analysis one can decide which 
energy out of various energies near various peaks are suitable. However, we have chosen the neutrino energy near the second
oscillation peak for $^{110}_{50}$Sn nuclei, for baseline of length 250 Km as for energy corresponding
to first oscillation peak we have to consider higher  boost factor $\gamma $ around 900.  Depending on the energy chosen near a peak one can appropriately choose  the boost factor $\gamma$ on which the neutrino energy window as shown in equation (\ref{eq:ewindow})  as well as  $\nu_e$ flux as shown in equation (\ref{eq:flux}) depend. This has been illustrated in figure \ref{fig1}.

\section{  Experimental set-ups}
For doing the analysis we choose four different setups:\
\newline
{\bf Setup(a)}: The length of the baseline is taken to be 130 Km (CERN-Fr\'ejus baseline) and the boost factor $\gamma$ to be 500 for nuclei  $^{110}_{50}$Sn.
\newline
{\bf Setup(b)}: The length of the baseline is taken to be 250 Km  and the boost factor $\gamma$ to be 320 for nuclei  $^{110}_{50}$Sn.
\newline
{\bf Setup(c)}: The length of the baseline is taken to be 130 Km (CERN-Fr\'ejus baseline) and the boost factor $\gamma$ to be 26 for nuclei  $^{152}$Yb.
\newline
{\bf Setup(d)}: The length of the baseline is taken to be 250 Km  and the boost factor $\gamma$ to be 40 for nuclei  $^{152}$Yb.

We consider a Water Cherenkov detector of fiducial mass 500 kt. Following \cite{signal}, the signal efficiency is considered to be 0.55 for  $\nu_\mu$ appearance channel. Background rejection factor coming from neutral current events is considered to be $10^{-4}$ for $\nu_\mu$ appearance channel. Signal error of $2.5\%$ and background error of $5\%$ has been considered.
For quasi-elastic $\nu_\mu$ appearance and $\nu_e$ disappearance we have followed signal efficiency and error as given in reference \cite{signal}.
We have considered the neutrino energy resolution as discussed earlier in (\ref{eq:ereso}) which can be obtained from vertex resolution after taking into account beam spreading.  The neutrino energy is known from
\eqref{eq:erest} and the energy width considered by us is obtained from \eqref{eq:ewindow} by considering the radius of the detector $R_{max}=100$m. We assume $10^{18}$ electron capture decays per year and the running time of  10 years for accumulating data. 

 We assume $10^{18}$ of $^{110}_{50}$Sn ions for boost. However, depending on the half life of Sn (4.11 hrs), the number of useful decays per effective  year ($10^{7}$ seconds ) considered are about
$0.608 \times 10^{18}$ with  boost factor ($\gamma = 500$) for    130 Km baseline and $0.768 \times 10^{18}$ with boost factor ($\gamma = 320$) for 250 Km baseline. Also we assume $10^{18}$ of $^{152}$Yb for boost. However, depending on the half life of $^{152}$Yb (3.04 seconds) and the boost factor $\gamma = 26$ or $\gamma = 40$, for baselines 130 Km or 250 Km respectively, the number of useful decays per effective  year ($10^{7}$ seconds ) considered are almost equal to the total number of nuclei i.e, $10^{18}$ as half life is  much smaller than  $^{110}_{50}$Sn. 
It is possible to achieve $\gamma$ about  480  at upgraded SPS facility at CERN \cite{sps,gam440,sps1} and $\gamma > 1000$ for LHC based design \cite{sps2}.
We have considered six energy bins keeping in mind the available energy window for different set-ups and the corresponding energy resolution in equation (\ref{eq:ereso}). In considering the energy resolution we have taken into account beam spreading. For that the the energy resolution considered by us is bad in comparison to the
energy resolution considered in reference \cite{rolinec} and we have to consider much lesser number of energy bins.  

We have compared discovery potentials of $CP$ violation obtained using monoenergetic neutrino beam to that using neutrino superbeam (SPL) facility at CERN. For that we have considered fifth setup as follows:
\newline
{\bf Setup(e)}: The length of the baseline is taken to be 130 Km corresponding to CERN-Fr\'ejus baseline.

For setup (e) we have considered water Cherenkov detector with fiducial mass of 500 kt, running time ($\nu + \bar{\nu}$) for 5+5 yrs. We have
considered beam intensity 2.4 MW and systematics on signal and background as 2\%. For other
detector characteristics like the signal efficiency, background rejection factor, migration matrices etc., we have followed  reference \cite{augustino}.

 We have considered the following central values and the priors for the oscillation parameters as mentioned in reference  \cite{1307.0807}: 
$|\Delta m^2_{31}|=2.5 \times 10^{-3}$, $\Delta m^2_{21}=7.5 \times 10^{-5}$, $\theta_{23}=38.3\degree$, $ \sin^2 \theta_{12}=0.31$, $\sin^2 2\theta_{13}=0.094$. Also we have considered prior of $5\%$ for $\sin^2 \theta_{12}$, $3\%$ for $\Delta m^2_{21}$, $8\%$ for $\theta_{23}$, $3\%$ for $\Delta m^2_{31}$ and $5\%$ for $\sin^2 2\theta_{13}$. A $2\%$ uncertainty is considered on the matter density.  As  the number of events corresponding to all set-ups  are quite large, any background due to atmospheric neutrinos are expected to be quite small and we have not considered such background in our analysis.

\section{Discovery reach of $CP$ violation}

Here we discuss the discovery of $CP$ violation for four different experimental set-ups (a-d)
mentioned earlier for monoenergetic neutrino beam. For comparison we also have presented 
the $CP$ violation discovery reach for superbeam as mentioned in experimental set-up (e).

In presenting our analysis we have considered the true hierarchy as normal hierarchy. However, we have  considered the uncertainty in the hierarchy of neutrino masses in the test values as it is not known
at present. For finding $CP$ violation we have fixed $\delta$(test) at $CP$ conserving $\delta$ values (0,$\pi$).  

\begin{figure}[H]
\centering 
\begin{tabular}{cc}
\includegraphics[width=0.4\textwidth]{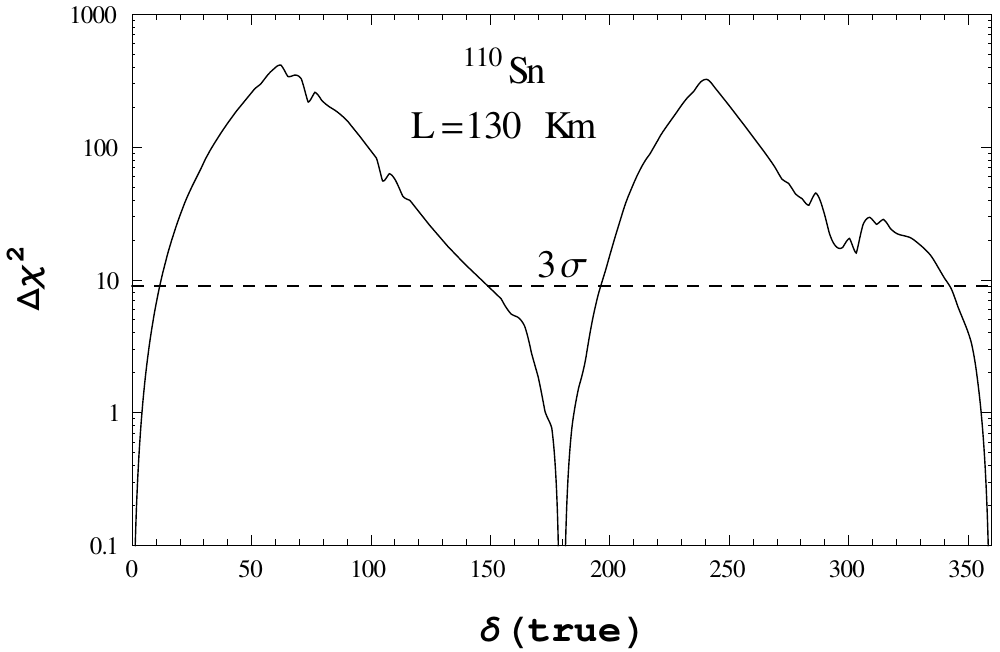}&
\includegraphics[width=0.4\textwidth]{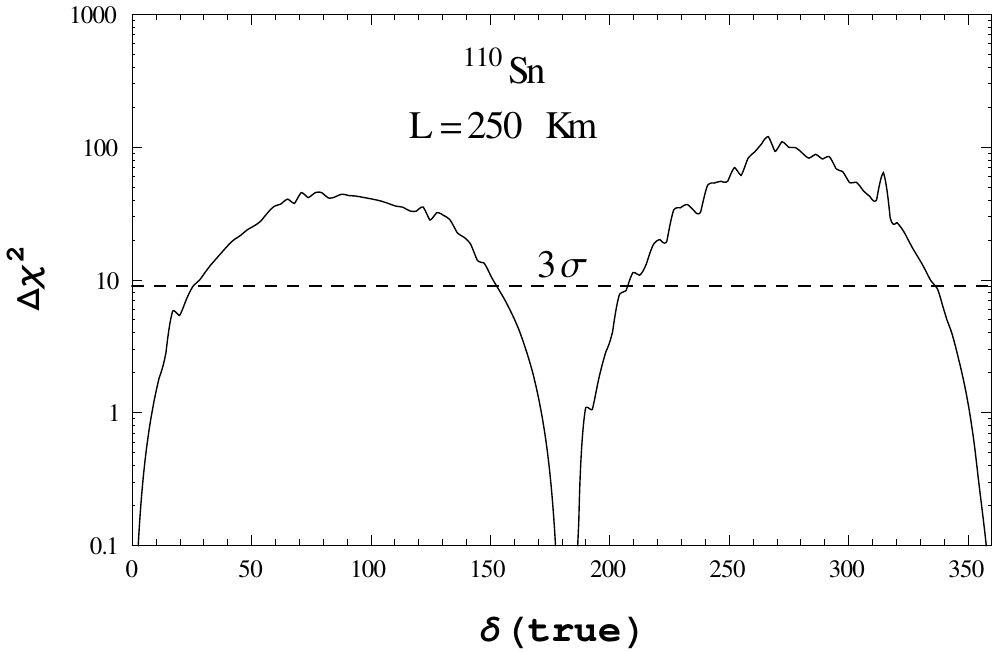}
\end{tabular}
\caption[] {{\small $\Delta \chi^2$ versus $\delta (true)$ for  two experimental set-ups 
(a) \& (b) with nuclei $^{110}_{50}$Sn.}}
\label{fig2}
\end{figure}

\begin{figure}[H]
\centering
\begin{tabular}{cc}
\includegraphics[width=0.4\textwidth]{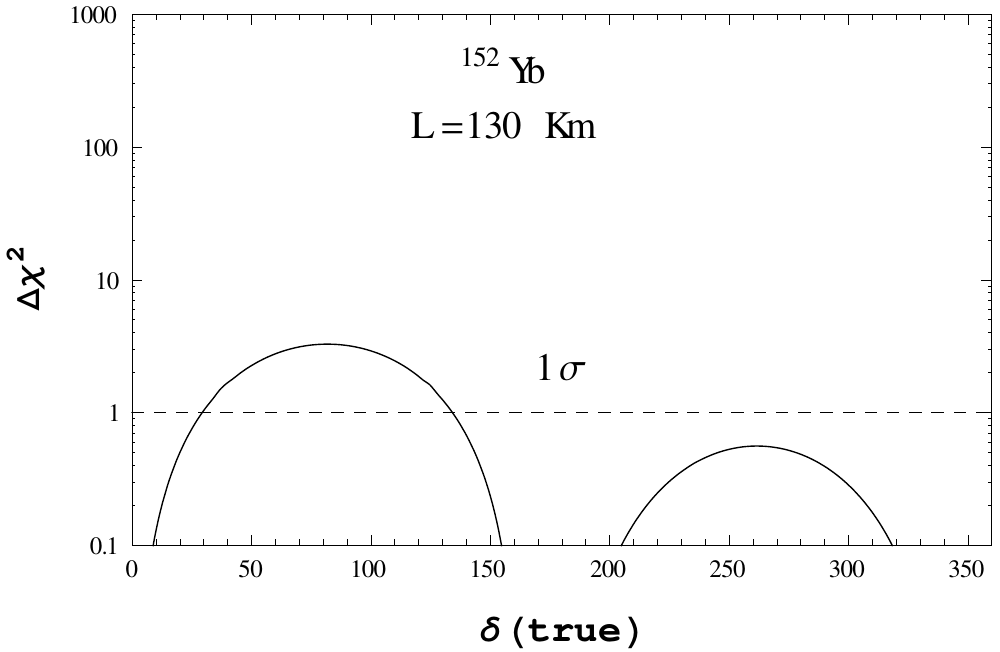}&
\includegraphics[width=0.4\textwidth]{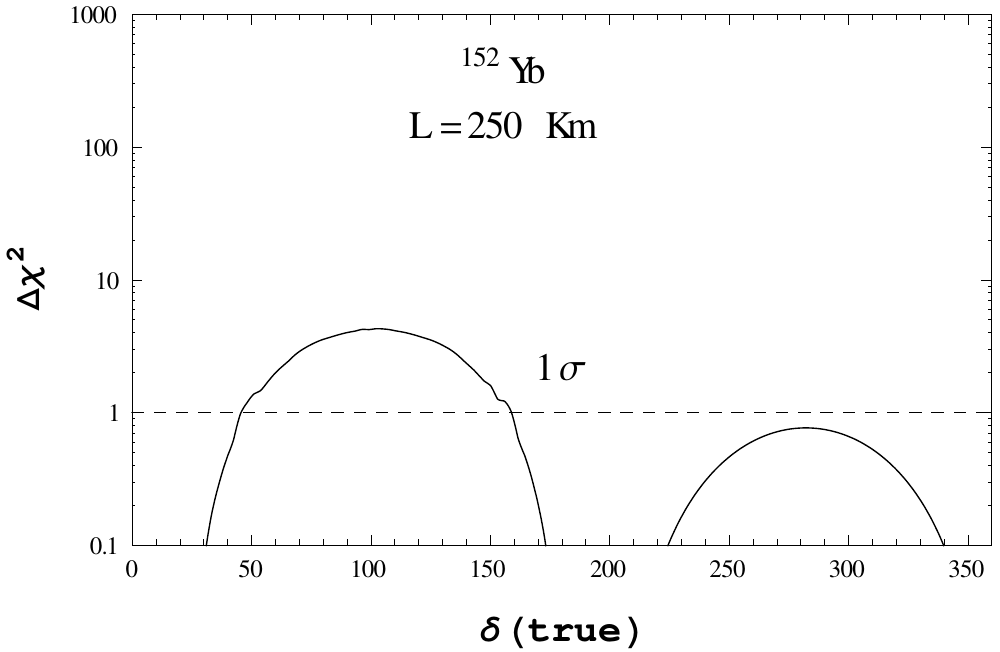}
\end{tabular}
\caption[] {{\small $\Delta \chi^2$ versus $\delta (true)$ for  two  experimental set-ups 
(c) \& (d) with nuclei $^{152}$Yb.}}
\label{fig3}
\end{figure}

In figure \ref{fig2}, $\Delta \chi^2$ versus $\delta (true)$ has been plotted to show the discovery reach  of the $CP$ violation for two different setups - setup(a) and setup(b) for $^{110}_{50}$Sn nuclei.  We find that the discovery of $CP$ violation for setup(a)  \&  (b)  
could be found for about 80.7\% and 72.7\% respectively of  the possible $\delta $ values
at $3\sigma$ confidence level.
In figure \ref{fig3},  $\Delta \chi^2$ versus $\delta (true)$ has been plotted to show the discovery reach  of the $CP$ violation for two different setups - setup(c) and setup(d) for $^{152}$Yb nuclei.  We find that the discovery of $CP$ violation for setup(c) and set-up (d) 
are not that good
and could be found for about 29.5\% and 31.8 \% at $1\sigma$ confidence level. For longer than 250 Km baselines we have not
presented any plots for $CP$ violation discovery reach. It seems one of the basic problem
for longer baselines will be relatively bad energy resolution because we are trying to use vertex resolution for getting energy resolution but there is beam spreading and as such over longer baseline beam spreading will make the energy resolution poorer. Instead of poorer energy resolution if we consider the same energy resolution which has been considered for 130 Km
then  for 650 Km baseline there could be $CP$ violation discovery reach of about 81.3 \% for $^{110}_{50}$Sn and 44.3\% for $^{152}$Yb. We have not presented any plots for this baseline because the energy resolution has been over-estimated. 


\begin{figure}[H]
\centering
\includegraphics[width=0.4\textwidth]{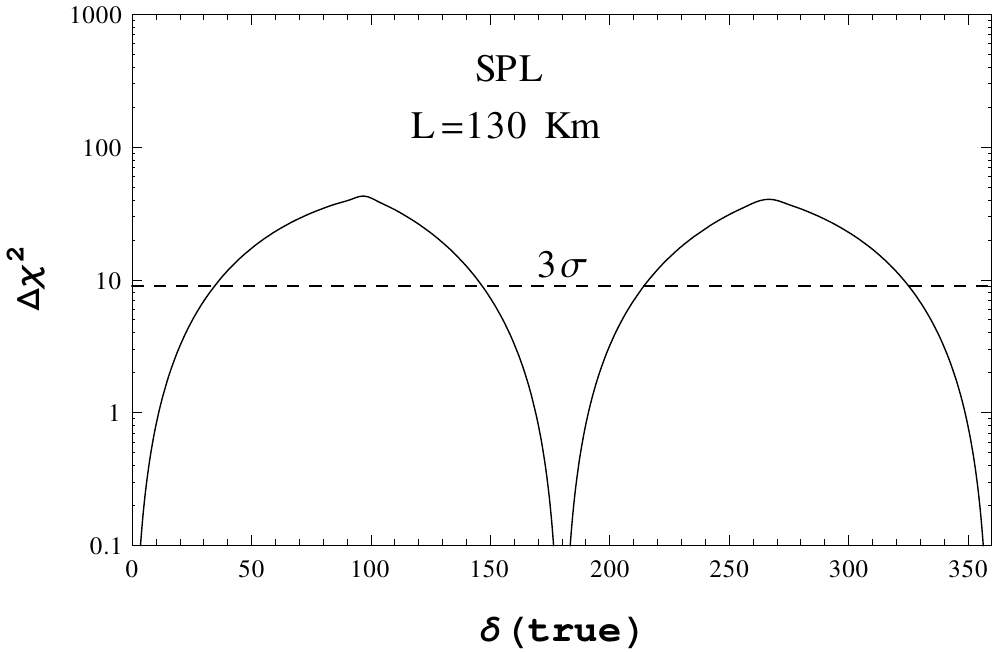}
\caption[] {{\small $\Delta \chi^2$ versus $\delta (true)$ for  130 Km baseline for superbeam .}}
\label{fig4}
\end{figure}
 
In our analysis we have chosen neutrino energy near the oscillation peak (as shown in figure \ref{fig1}) which is more $\delta$ sensitive region. This consideration improves the $CP$ violation discovery reach.
In \cite{rolinec} (as shown in figure 7 of that paper) $CP$ violation discovery reach has been shown to be  about 81\% of the possible $\delta $ values for their setup II for 250 Km baseline for the presently known
$\theta_{13}$ value. The discovery reach seems to be better than that we have presented here. This is because $\gamma$ value considered there was large and also beam spreading has not been taken into account 
for considering energy resolution at the detector. In our analysis we have considered more realistic value of $\gamma$ which could be achievable at present (keeping in mind the possible SPS upgrade at CERN). We have also taken into account the beam spreading in our present analysis in estimating  effective energy resolution and as such it is not as good as considered in reference \cite{rolinec} and for that we have taken only a few neutrino energy bins for the analysis unlike \cite{rolinec}.  Due
to lesser $\gamma$ value the flux is also reduced with respect to that considered in reference 
\cite{rolinec}. Furthermore, particularly for $^{110}_{50}$Sn isotopes, the number of useful decays which we have considered is smaller by 0.608 and 0.768 times the number  considered in that reference for baselines 130 Km and 250 Km respectively as we have considered actual decays of this ion in one year.     

In figure \ref{fig4}, $\Delta \chi^2$ versus $\delta (true)$ has been plotted to show the discovery reach  of the $CP$ violation for experimental setup (e) for superbeam. We find that the discovery of $CP$ violation 
could be found for about 63.6\%  of  the possible $\delta $ values
at $3\sigma$ confidence level. Here for comparison we have considered same baseline of 130 Km. 
It seems for shorter baseline of 130 km experimental set-up with monoenergetic beam could have better discovery reach for $CP$ violation in the leptonic sector than that with superbeam facility.

There are several works \cite{others}
for finding the $CP$ violation discovery reach in the context of superbeam, neutrino factory  and beta beam. Recent studies for such neutrino sources show that $CP$ violation discovery reach could be as good as about 72 \% of  the possible $\delta $ values
at $3\sigma$ confidence level for superbeam \cite{superbeam} for 500 Km baseline, 85 \% of  the possible $\delta $ values
at $5\sigma$ confidence level for neutrino factory \cite{nufact} for 2000 Km baseline and 70 \% of  the possible $\delta $ values
at $3\sigma$ confidence level for beta beam  \cite{betabeam} for 650 Km baseline. So irrespective of baselines, if we compare the $CP$ violation discovery reach for different neutrino sources, then monoenergetic neutrino beam could be better than superbeam or beta beam facilities, but in neutrino factory the $CP$ violation discovery reach could be even better.

\section{Conclusion}
For a comparative study we have considered two different type of monoenergetic neutrino sources to find the discovery reach of the $CP$ violation - one   $\nu_e$ source is from electron capture  decays of $^{110}_{50}$Sn isotopes  and the other  $\nu_e$ source is from electron capture  decays of  isotopes$^{152}$Yb.
For each case we have considered two baselines 130 Km and 250 Km. For experimental set-ups 
(a-d)  water Cherenkov detector has been considered. For comparison we have considered 130 Km baseline with experimental set-up (e) where both neutrino and anti-neutrino sources from superbeam has been considered. 

When one considers technical issues involved in the accelerator and running the ions through
vacuum tube, isotopes$^{152}$Yb  is better candidate than  $^{110}_{50}$Sn isotopes because of
much lesser half life.  $^{152}$Yb is also better because of the  dominant electron capture decay to one energy level. However, as can be seen from figures the discovery reach of $CP$ violation is found to be better for  $^{110}_{50}$Sn isotopes because of the scope to consider higher $\gamma $ values for shorter baselines giving larger neutrino flux. For $^{152}$Yb one could consider higher $\gamma $ values
for longer baselines but  due to matter effect and more neutrino flux suppression and not so good energy resolution that is not a good option for $CP$ violation discovery. Out of different baselines for $^{110}_{50}$Sn nuclei,  we find slightly better discovery reach for shorter baseline of 130 Km with
$\gamma = 500$. 

Following the recent analysis of $CP$ violation discovery reach it is found that the best discovery reach could be possible for neutrino factory. However, the discovery reach with monoenergetic beam could be better than superbeam and beta beam.  
This  is  primarily due to the scope of almost precise knowledge  of neutrino  energy  and also for 
the presence of purely one type of neutrino flavor ($\nu_e$) in the beam \cite{rolinec} and also for the scope of adjusting the
almost monoenergetic neutrino energy to a suitable  oscillation peak in the oscillation probability $\nu_e \rightarrow \nu_\mu$
where variation of probability due to $\delta $ is more significant and the probability is relatively higher.
  
After the discovery of non-zero neutrino vacuum mixing angle
$\theta_{13}$ by Daya Bay  \cite{daya}, RENO \cite{reno}, Double Chooz \cite{double} collaborations and also by other experiments it is now very important to know whether there is $CP$ violation in the leptonic sector and  $\delta $ has some different value from $(0, \pm \pi)$. For this it seems that considering the neutrino source  from electron capture decays could be quite worthwhile in future. However, building up of such facility of neutrino beam will require some technological development and the implementation of it might take some time \cite{Lind}. The existing CERN accelerator complex could be used to study such facility. However, the monoenergetic neutrino flux require a very large number of ions to be stored in the decay ring. It is difficult to control the beam at high intensities because of space charge detuning, intra beam scattering and vacuum loss.  With SPS upgrade it could be possible to accelerate the ions to $\gamma =480$ but accelerating above that seems difficult \cite{sps,gam440,sps1}. Depending on the half life
of $^{110}_{50}$Sn we have reduced the total number of useful decays of the ion per effective year from $10^{18}$ but the value considered is
still extreme because of the requirement of acceleration and storage of the partially charged ion. For improving this the vacuum conditions in
SPS would be required to be upgraded.  It requires more study on such beam facility. With technological improvement if it is possible to consider monoenergetic beam with $\gamma > 1000$
\cite{sps2}, then the $CP$ violation discovery reach will improve further than what has been
presented in this work.

\hspace*{\fill}

\noindent
{\bf  Acknowledgment:}  Z. R likes to thank
University Grants Commission, Govt. of India for providing research fellowship. Z. R and R. A thank A. Dasgupta for discussion.

\end{document}